# Data Mining of the Concept «End of the World» in Twitter Microblogs

Bohdan Pavlyshenko

*Ivan Franko Lviv National University, Ukraine, pavlsh@yahoo.com*

**Summary**

This paper describes the analysis of quantitative characteristics of frequent sets and association rules in the posts of Twitter microblogs, related to the discussion of "end of the world", which was allegedly predicted on December 21, 2012 due to the Mayan calendar. Discovered frequent sets and association rules characterize semantic relations between the concepts of analyzed subjects. The support for some fequent sets reaches the global maximum before the expected event with some time delay. Such frequent sets may be considered as predictive markers that characterize the significance of expected events for blogosphere users. It was shown that time dynamics of confidence of some revealed association rules can also have predictive characteristics. Exceeding a certain threshold, it may be a signal for the corresponding reaction in the society during the time interval between the maximum and probable coming of an event.

## Introduction

The system of microblogs Twitter is one of popular means of interaction among users via short messages (up to 140 characters). Twitter messages are characterized by high density of contextually meaningful keywords. This feature conditions the availability of the study of microblogs by using data mining in order to detect semantic relationships between the main concepts and discussion subjects in microblogs. Very promising is the analysis of predictive ability of time dependences of key quantitative characteristics of thematic concepts in the messages of twitter microblogs.

As it is known, the end of the world, which was supposedly to take place on December 21, 2012, according to the Mayan calendar, has been discussed intensively in the blogosphere. Obviously, there were no scientific prerequisites for this prediction. Even NASA has provided a clarification on this matter (http://www.nasa.gov/topics/earth/features/2012.html). However, the very stream of information, which is associated with the "end of the world" concept, may have an impact both on individuals and on the entire society. Certain combinations of semantic concepts, which were periodically repeated from different information sources, can play a role of information viruses and affect people's mentality and behavior, cause certain trends in society. Therefore, in our opinion, the development of methods for detecting trends in the information stream is very promising. The discussion of notions related to the concept "end of the world" has a number of special features. These features form the dynamics of information flow, associated with the discussion of the concept "end of the world ".

The peculiarities of social networks and users' behavior are researched in many studies. In [Java, 2007; Kwak, 2010] microblogging phenomena were investigated. In [Newman, 2003] it is shown that social networks differ structurally from other types of networks. User influence in twitter was studied in [Cha, 2010]. Users' behavior in social networks is analyzed in [Benvenut, 2009]. In [Pak, 2010] the methods of opinion mining of twitter corpus were analyzed. Several papers are devoted to the analysis of possible forecasting events by analyzing messages in microblogs. In [Bollen, 2011] it was studied whether public mood as measured from large-scale collection of tweets posted on twitter.com, is correlated or predictive for stock markets. In [Asur, 2010] it is showen that a simple model built from the rate, at which tweets are created about particular topics, can outperform market-based predictors. In [Mishne, 2006] films sales based on the discussions in microblogs are analyzed. In [Shamma, 2010] the twitter activity during media events was investigated. The paper [Gruhl, 2004] studied the phenomenon of resonance in blogospheres, which can be caused by events in real world. The analysis of search engines queries is also used in forecasting [Choi, 2009; Wu, 2010].

The discussion of "end of the world" topic on December 21, 2012, due to the Mayan calendar, has a number of peculiarities. Firstly, some bloggers have doubts about the outcome of events in the predicted date, some others are sure about negative consequences of events, but the majority is optimistic and does not believe in negative predictions.

In this paper we construct a set-theoretic model of key tags of twitter messages. Then we identify frequent itemsets of words in messages with the key tags {end, world}, construct association rules based on the detected frequent sets. We also consider the time dynamics of found frequent sets and association rules.

**Theoretical Model**

Let us consider a model that describes microblogs messages. W have choosen some set of keywords which specify the themes of messages and are present in all messages, for example, $kw \in Keywords$, $Keywords = \{end, world\}$. Then we define a set of microblogs messages for the analysis:

$$TW^{kw} = \{tw^{(kw)}{}_i \mid kw_j \in tw_i,\ kw_j \in Keywords\}. \tag{1}$$

Our next step is to consider the basic elements of the theory of frequent sets. Each tweet will be considered as a basket of key terms

$$tw_i = \{w_{ij}^{tw}\}. \tag{2}$$

Such a set is called a transaction. We label some set of terms as

$$F = \{w_j\}. \tag{3}$$

The set of tweets, which includes the set $F$ looks like

$$TW_F^{kw} = \{tw_r \mid F \in tw_r; r = 1,...m\}. \tag{4}$$

The ratio of the number of transactions, which include the set $F$, to the total number of transactions is called a support of $F$ basket and it is marked as $Supp(F)$:

$$Supp(F) = \frac{|TW_F^{kw}|}{|TW^{tw}|}. \tag{5}$$

A set is called as frequent, if its support value is more than the minimum support that is specified by a user

$$Supp(F) > Supp_{min}. \tag{6}$$

Given the condition (6) we find a set of frequent sets

$$L = \{F_j \mid Supp(F_j) > Supp_{min}\}. \tag{7}$$

For identifying frequent sets an Apriori algorithm [Agrawal, 1994; Mannila, 1994] is mainly used. It is based on the principle that the support of some frequent set does not exceed the support of any of its subsets. Based on frequent sets we can build association rules, which are considered as

$$X \rightarrow Y, \tag{8}$$

where $X$ is called *antecedent* and $Y$ is called *consequent*. The objects of *antecedent* and *consequent* are the subsets of the frequent set $F$ of considered keywords

$$X \cup Y = F. \tag{9}$$

In the problems of finding association rules two major phases are distinguished: the search of all frequent sets of objects and the generation of association rules based on detected frequent sets. Using frequent set, one can build a large number of association rules, which will be defined by different combinations of features. For the evaluation and selection of useful rules a number of quantitative characteristics is introduced, in particular *support* and *confidence*. *The support* of an association rule shows what part of transactions contains this rule. Since the rule is based on the frequent set of considered keywords, the rule $X \to Y$ has the support equal to the support of the set $F: X \in F, Y \in F$. Different rules, based on the same set, have the same support values. The support is calculated using the formula (5). The confidence of the association rule shows the probability of the fact that the presence of the subset of X attribute in the transaction implies the presence of the subset of Y attribute. Confidence is defined as the ratio of transactions containing the subsets X and Y attributes to the number of transactions containing the subset of X attribute only:

$$Conf_{X \to Y} = \frac{\left|TW_{X \cup Y}^{kw}\right|}{\left|TW_X^{kw}\right|} = \frac{Supp_{X \cup Y}}{Supp_X}. \qquad (10)$$

An important feature is that different association rules of one and the same set will have different confidence.

**Experimental Part**

To implement the experimental research a package of applied programs in Perl language has been developed. With the help of this package and using the API of Twitter system a test array of messages containing the key words {end, world} was downloaded. The messages were being loaded periodically (every 30 seconds) since December 10, 2012 to January 5, 2013. We have analyzed the daily dynamics of such characteristics of association rules as support and confidence. The total amount of messages under analysis is 987,000. The array of posts obtained is a result both of bloggers' own thoughts and the discussions and changing of participants' viewpoints in these discussions. From the downloaded data array we have removed the stop-words of high frequency and random and rarely used words of low-frequency, which had met less than 10 times in the array of messages. On the basis of the array obtained the search of frequent sets of keywords was found. Semantically meaningful sets, regarding the themes under analysis, were selected from the obtained array of frequent sets. Here are some examples of frequent sets of key words, the support of which is more than 300:

*{annoying, day, 21st}, {annoying, december, 21st}, {anything, finish, mayans},*
*{anything, survived}, {believes, drop, everyone}, {calendar, mayan}, {die, friday, all},*
*{era, new, mayans, december}, {explains, nasa}, {finish, calendar, mayan}, {bad, thing},*
*{magic, december, 21st}, {mayans, december, 21st}, {must, happen, this},*
*{new, mayans, december}, {sacrificed, themselves, mayan}, {wrong, worry, mayans}.*

The frequent sets obtained are the basis for the construction of association rules $X \to Y$. To generate association rules let us find the subsets of frequent sets that have a high level of support. Such subsets can be considered as antecedents for association rules. The words like *21st, friday, mayan* may be regarded as the examples of such antecedents. Given that the messages containing the words {end, world} are under consideration, then the antecedents of association rules will look like *{end, world, 21st}, {end, world, friday}, {end, world, mayan}*. Based on the identified frequent sets we have constructed the association rules that reflect the semantics of the relationships of thematic terms. Table 1 shows some constructed association rules. The table shows the characteristics of the support $Supp(F)$ and confidence for the association rules. The support $Conf_{X \to Y}$ describes the rule of $X \to Y$ type, and the support $Conf_{Y \to X}$ describes the rule of $Y \to X$ type.

*Table 1. Association rules based on the frequent sets of semantic features.*

| X | Y | $Supp(F)$ | $Conf_{X \to Y}$ | $Conf_{Y \to X}$ |
|---|---|---|---|---|
| 21st december | mayans | 0.000700 | 0.0501 | 0.0223 |
| 21st december | survived | $4.9610*10^{-5}$ | 0.0035 | 0.0035 |
| 21st december | worry | $9.4157*10^{-5}$ | 0.0067 | 0.0073 |
| 21st december | apocalypse | $3.7460*10^{-5}$ | 0.0026 | 0.0068 |
| 21st december | friday | 0.0003868 | 0.0277 | 0.0099 |
| 21st december | die | 0.0001286 | 0.0092 | 0.0117 |
| 21st december | survive | $4.9610*10^{-5}$ | 0.0020 | 0.0107 |
| 21st | die | 0.0007462 | 0.0161 | 0.0682 |
| 21st | worry | 0.0012585 | 0.0272 | 0.0981 |
| 21st | apocalypse | 0.0001934 | 0.0041 | 0.0352 |
| 21st | mayans | 0.0006237 | 0.0303 | 0.0447 |
| friday | worry | 0.0001691 | 0.0043 | 0.0131 |
| friday | die | 0.0011674 | 0.0300 | 0.1068 |
| friday | apocalypse | 0.0001569 | 0.0040 | 0.0285 |
| friday | mayan | 0.0003766 | 0.0160 | 0.0328 |
| mayan | worry | $8.3021*10^{-5}$ | 0.0043 | 0.0064 |
| people | believe | 0.0029250 | 0.0675 | 0.1303 |
| people | survived | 0.0002025 | 0.0046 | 0.0145 |
| people | worry | 0.0001093 | 0.0025 | 0.0085 |
| people | annoying | 0.0002622 | 0.0060 | 0.1026 |
| people | stupid | 0.0017100 | 0.0394 | 0.2586 |

Some rules are trivial and obvious, the others characterize the tendencies of analyzed themes. Analysing the dynamics of the values $Supp(F)$, $Conf_{X \to Y}$, $Conf_{Y \to X}$, one can define the trends of discussion topics in microblogs. It is obvious, that only the association rules with the highest value of confidence deserve for further consideration. Given the large amount of posts array and the dictionary word composition of this array, the values of confidence for most association rules will be small. High values of confidence may occur in spam and trivial association rules that bear no new semantic component for the array under analysis. Obtained association rules show the semantic network of concept links that characterize the subject area of discussion in the analyzed array of messages. Since the main part of discussion is a time guide (December 21, 2012), we study the time dynamics of the quantitative characteristics of frequent sets and association rules. We will also calculate the time dependences of support values for typical key terms and frequent sets.

While analyzing the dynamics of frequent sets of keywords, one may single out the following groups due to the maximum :
1) frequent sets with the maximum of 7-8 days before the expected event on December 21,2012;
2) frequent sets with the maximum of 1-3 days before December 21,2012;
3) frequent sets with the maximum exactly on December 21,2012;
4) frequent sets with the maximum after December 21,2012;
5) frequent set of periodic oscillations;

Figures 1-5 show the time dynamics of frequent sets of these types. Obviously, the most interesting are the frequent sets with their maximum just before the expected event. Such sets of key words can be considered as predictive markers. By the nature of their maximum one can judge to what extent the discussion of an expected event may be socially meaningful. "An event" in terms of concept under analysis is a date of receiving an answer to the discussed forecast. The presence of time delay between the maximum of characteristics and the date of expected event gives the opportunity to react. For example, in case of necessity, considering prognostic markers, the government may resort to follow-up actions, within the time delay period between the maximum characteristics and the expected date. Such actions may be, for instance, additional explanations, as it was done by NASA, since it has been known beforehand that there were no scientific prerequisites for anxiety. Yet for a certain part of the population the topic of "end of the world" on December 21, 2012 caused serious concern.

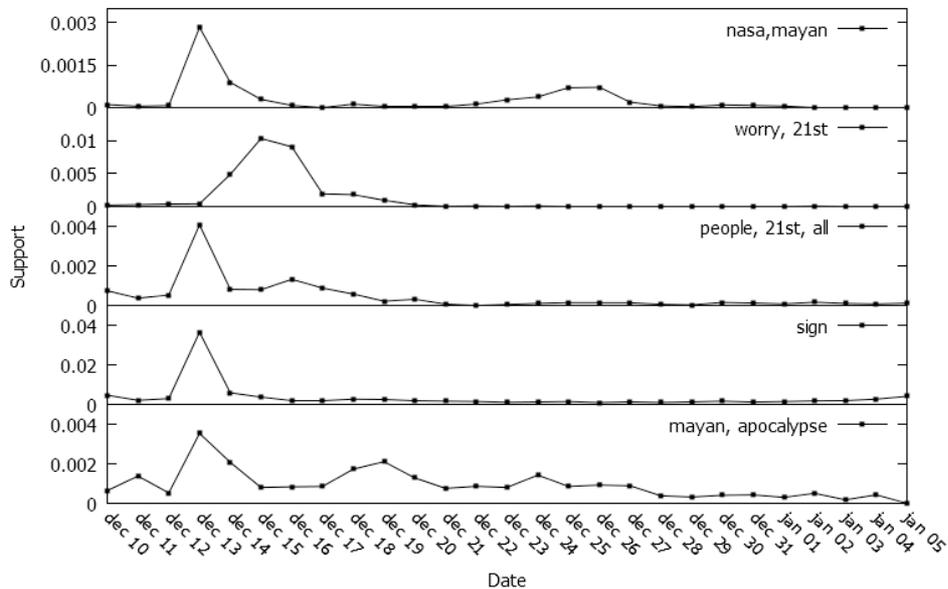

*Fig.1 The dynamics of frequent sets support with the maximum of 6-8 days before the predicted date.*

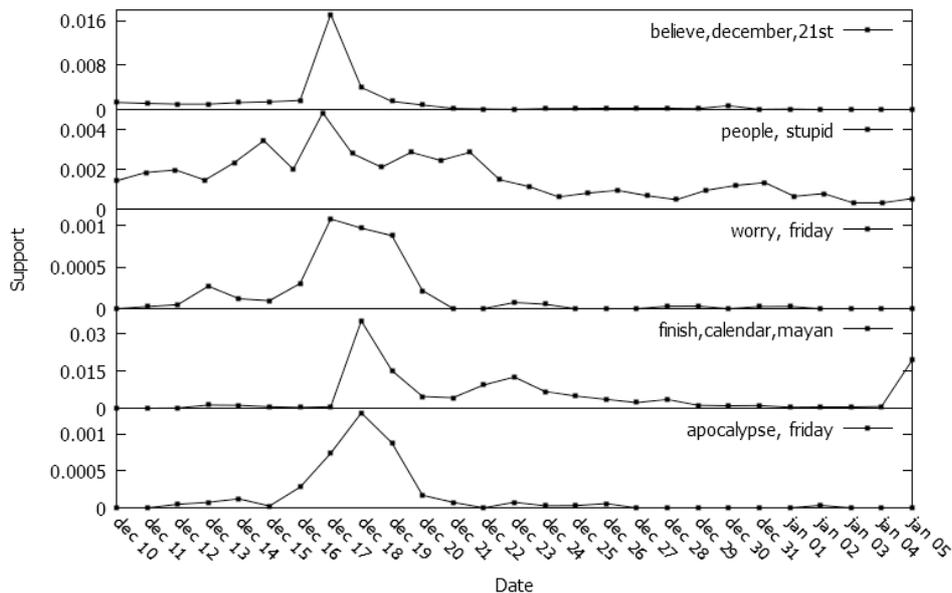

*Fig.2 The dynamics of frequent sets support with the maximum of 3-5 days before the predicted date.*

Here are the examples of revealed key words and frequent sets with global maximum before the expected event:

Maximum on Dec 13: *sign, {21st,apocalypse}, {explain,nasa}, {mayan,apocalypse}, {nasa,mayan}, {people, 21st,all};*

Maximum on Dec 15: *21st, soon, {worry, 21st };*

Maximum on Dec 17: {worry, *friday}, {people,stupid}, {believe, december, 21st };*

Maximum on Dec 18: *plague, people, finish, {mayan, friday}, {mayan, calendar}, {finish,calendar,mayan};*

Maximum on Dec 19: *{people, friday}*;
Maximum on Dec 20: *tomorrow*.

Some frequent sets have several peaks before the expected date of forecast. For example, for the set *{21st, friday}* (Fig. 6) there is a maximum on December 13 and 18. Figure 6 shows the support and the confidence of association rules *{apocalypse => 21st}*, *{21st => friday}*, *{mayan => friday}*. These association rules also have predictive potential, since they reach the maximum support and confidence before the expected date. The quantitative value of association rule confidence can be considered as an additional predicting factor along with support.

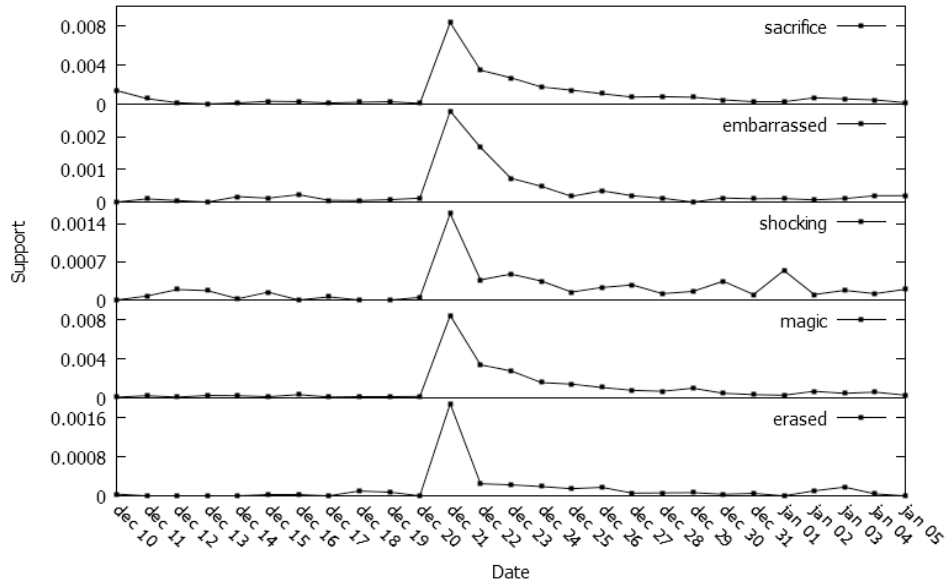

*Fig. 3 The dynamics of frequent sets support with the maximum on the predicted date.*

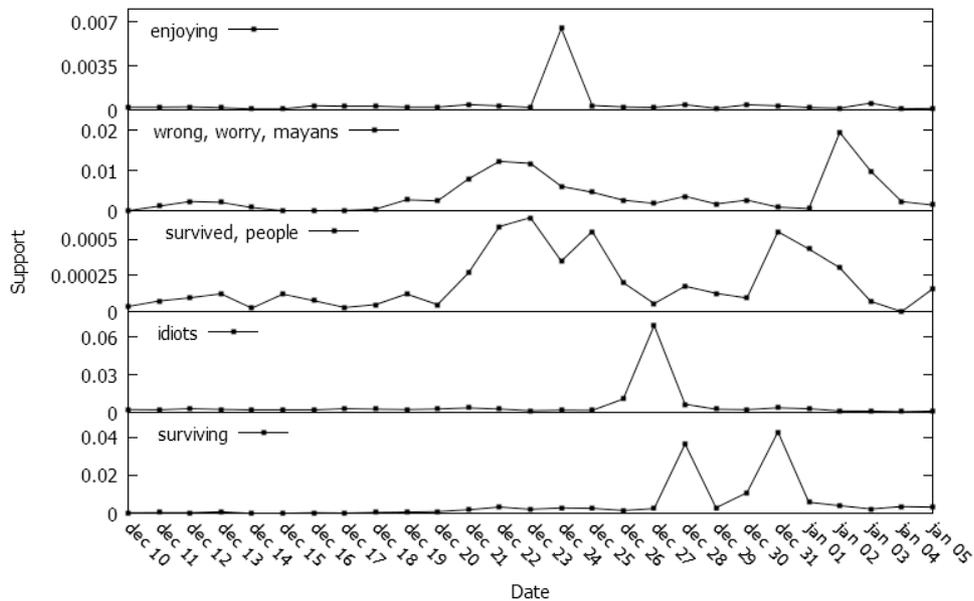

*Fig. 4 The dynamics of frequent sets support with the maximum after the predicted date.*

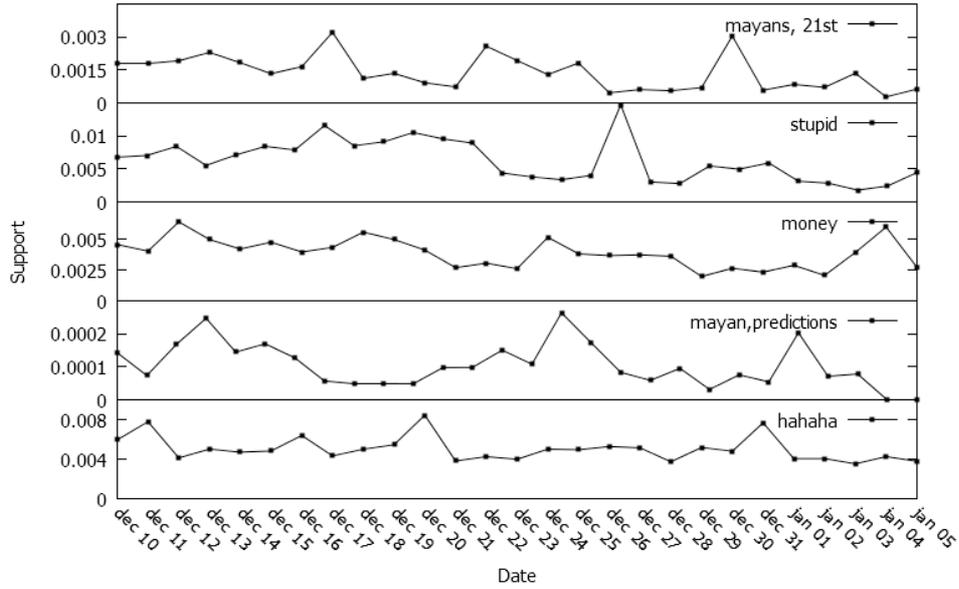

*Fig.5 The dynamics of frequent sets support with periodic changes
in the analyzed time interval.*

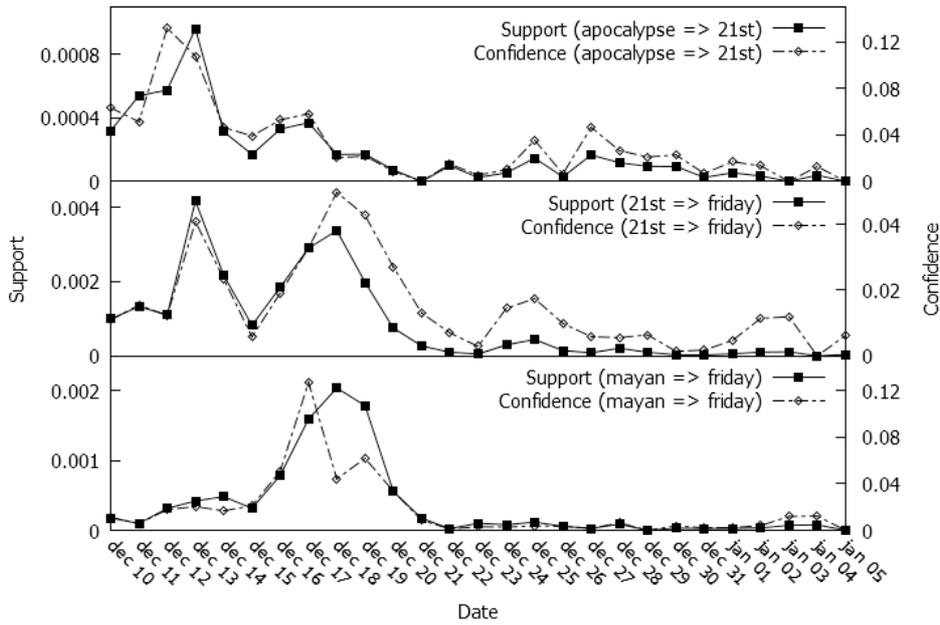

*Fig. 6 The dynamics of support and confidence of association rules
with the predictive potential.*

We have also found the sets which have the maximum before and after the date of the forecast. For example, Fig. 7 shows that for *Supp (mayans, 21st, december), Conf (mayans => {21st, december}), Conf ({21st, december} => mayans)* there are maxima on December 17 and 22, and minimum on the expected date December 21.

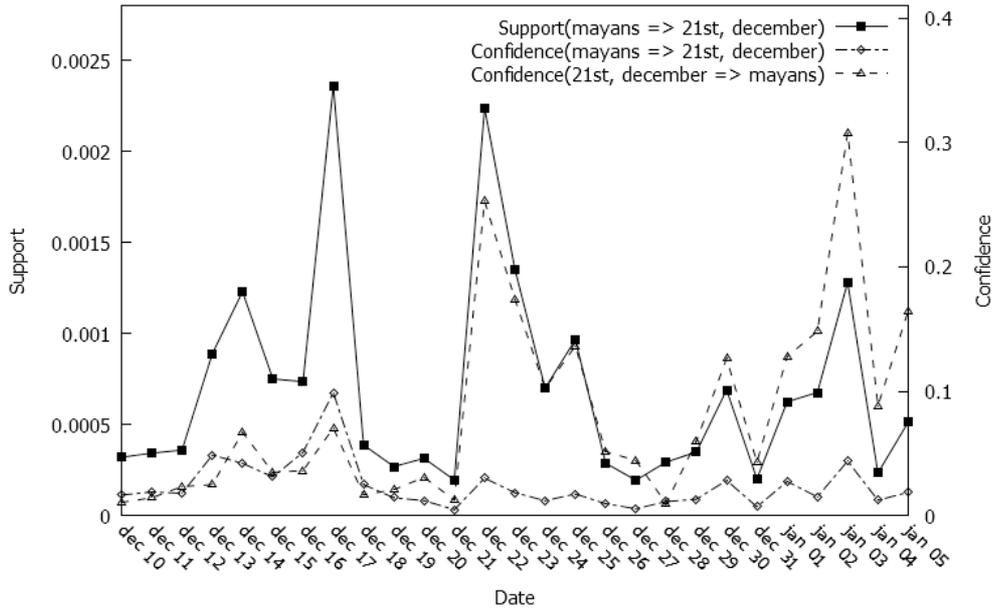

*Fig. 7 The dynamics of support and confidence of the asssociative rule
with the maximum around the time guide of the expected date of forecast.*

The frequent sets with the maximum after the expected prediction (Fig. 5) characterize the discussion of the forecast results that did not prove to be correct. The type of frequent sets with periodical changing frequency (Fig. 5) may characterize the discussion the new turns of which arise on the basis of intermediate conclusions and decisions.

**Discussion and conclusions**

The results obtained show that the discussion of "end of the world" on December 21, 2012, forecasted on the basis of the Mayan calendar, can be characterized by the dynamics of identified key words and frequent sets, associated with the concept "End of the World ". Discovered association rules characterize semantic relations between the concepts of analyzed subjects. The dynamics of support and confidence of some of the identified association rules reflect information trends in discussing the forecast under analysis and the expected event. The event in the analyzed context is the coming of the date of getting answers to the announced forecasts. One can select a subset of frequent sets which get the maximum support before the expected event (December 21, 2012) with given time interval between the maximum and the event, expected by the participants of the blogosphere. Such frequent sets can be considered as potential predictive markers that can characterize the significance of an event under discussion for blogosphere users. If these parameters exceed a certain threshold, it may be a signal for the corresponding reaction in the society during the time interval between the maximum and probable coming of an event. Such a reaction can be, for example, additional explanations, as it was done by NASA. The markers of analyzed events have the prognostic potential both for informational flows and for the events that may occur as a result of impact of these information flows on the participants of a blogosphere, and their making appropriate decisions. The results obtained show that some topics of discussions can affect bloggers community, cause periodic informational reactions and predictable information flows. Along with frequent sets, the quantitative characteristics of association rules may also have prognostic properties. In particular, there is a predictive global maximum in the dynamics of support and confidence found in some association rules. The frequent sets the support of which is minimal in the analyzed time guide, characterize bloggers' topics who are currently in the standby process and stopped the discussion. This information process is somewhat similar to a viral epidemic, it is a new characteristic property of social networks, with the absence of which the scales of discussion of unscientific topics would be much lower. Obviously, some of the identified keywords and frequent sets with the maximum before the expected event are situational and have no predictive potential. They may be caused by the activity of some random participants of discussions on a given topic. Therefore in order to identify the most characteristic predictive markers characterizing the significance of analyzed forecast or expected event for the participants of a blogosphere, it is essential to continue monitoring similar reports in the blogs, where nonscientific forecasts are being discussed. One of the ways of singling out the predictive frequent sets of key terms and removing casual frequent sets, is the analysis whether the words of the frequent sets under study belong to given thematic field. A thematic field can be a predetermined set of words, which characterizes the topics under research. The frequent sets of words, that are not included into thematic field, may be discarded as situational ones and not specific for analyzed topics. In our further research we are planning to explore the methods of automated detection of prognostic markers.